\documentstyle[epsfig,aps,prb,twocolumn,floats]{revtex}

\begin{document}

\newcommand{\gsim}{
\,\raisebox{0.35ex}{$>$}
\hspace{-1.7ex}\raisebox{-0.65ex}{$\sim$}\,
}

\newcommand{\lsim}{
\,\raisebox{0.35ex}{$<$}
\hspace{-1.7ex}\raisebox{-0.65ex}{$\sim$}\,
}

\newcommand{\const}{ {\rm const} }
\newcommand{\arctanh}{ {\rm arctanh} }

\bibliographystyle{prsty}

\title{ \begin{flushleft}
%{\small \em submitted to}\\
{\small 
PHYSICAL REVIEW B $\qquad\qquad$
\hfill
VOLUME {\normalsize 59}, 
NUMBER {\normalsize 5} 
\hfill 
%MONTH XXX
{\normalsize 1} FEBRUARY {\normalsize 1999}-I, 
{\normalsize 3671-3674}
}\\
\end{flushleft}  
Quantum-classical escape-rate transition of a biaxial spin system with a
longitudinal field: A perturbative approach
}

\author{
D.~A. Garanin \cite{e-gar} 
}

\address{
Max-Planck-Institut f\"ur Physik komplexer Systeme, N\"othnitzer Strasse 38,
D-01187 Dresden, Germany\\ }

\author{
E. M. Chudnovsky \cite{e-chu}
}

\address{
Department of Physics and Astronomy, City University of New York--Lehman College, \\
Bedford Park Boulevard West, Bronx, New York 10468-1589 \\
%}
%\date{18 November 1997}
%\maketitle
%\abstract{
\smallskip
%{\rm (\today) }
{\rm (Received 4 August 1998) }
\bigskip\\
\parbox{14.2cm}
{\rm
The quantum-classical transition of the escape rate of the spin model 
${\cal H} = -DS_z^2 - H_zS_z + B S_x^2$ is investigated by a perturbative
approach with respect to $B$ [D. A. Garanin, J. Phys A {\bf 24}, L61 (1991)]. 
The transition is first order for $B<B_c(H_z)$, the boundary line going to
zero as $B_c/D \sim 1 - H_z/(2SD) $ in the strongly biased limit.
The range of the first-order transition is thus larger than for the model 
${\cal H} = -DS_z^2 - H_zS_z - H_x S_x$ studied earlier, where in the
strongly biased case $H_{xc}/(2SD) \sim [1 - H_z/(2SD)]^{3/2}$. 
The temperature of the  quantum-classical transition,  $T_0$, behaves linearly
in the strongly biased case for both models: $T_0 \sim 2SD -H_z$.
\smallskip
\begin{flushleft}
PACS number(s): 75.45.+j, 75.50.Tt
\end{flushleft}
} 
} 
\maketitle

\section{Introduction}

Recently we have shown \cite{garchu97,chugar97} that the order of the quantum-classical escape-rate
transition \cite{chu92} in the spin system described by the Hamiltonian 
${\cal H} = -DS_z^2  - H_x S_x$ is controlled by the transverse field $H_x$.
According to Ref. \onlinecite{chugar97}, for the reduced field $h_x \equiv H_x/(2SD)$ in the interval 
$1/4 \leq h_x < 1$ the transition is  second order, which is a common situation
(for $h_x>1$ the barrier between the two wells disappears).
In contrast, for $0<h_x < 1/4$ there is a first-order transition of the
escape rate $\Gamma(T)$ characterized by the discontinuity of $d\Gamma/dT$ at
the transition temperature $T_0$ in the large-spin limit.    
Subsequent calculations \cite{garmarchu98} rendered the whole phase diagram
for the model with an arbitrarily directed field,  ${\cal H} = -DS_z^2 -
H_zS_z - H_x S_x$, where the first-order transition occurs for $h_x$ below the
line $h_{xc}(h_z)$, where $h_z \equiv H_z/(2SD)$.
This line starts from the value 1/4 at $h_z=0$ and approaches zero as $h_{xc}
\sim (1-h_z)^{3/2}$ in the strongly biased limit.
This limit can be relevant for observation of the crossover between first- and
second-order transitions on single-domain magnetic particles, where the
barrier should be lowered by applying the field to make escape rates
measurable.
On the other hand, for molecular magnets, such as Mn$_{12}$ ($S=10$, $D=0.6$~K),
experiments can be done even in the unbiased case, $H_z=0$.

The mechanism leading to an exotic first-order escape-rate transition in spin
systems is the following.
In the absense of the transverse field the eigenstates of the system are those
of the operator $S_z$, and there is no tunneling between the wells.
This implies that the barrier of the effective potential of the spin system, which can be
obtained by a mapping onto a particle problem,  \cite{zas90pla,chugar97}
becomes infinitely thick in the limit $H_x \to 0$, preserving, however, its
height $\Delta U$ (see Fig. 1 of Ref. \onlinecite{garmarchu98}).
That is, the barrier has a very flat top and it resembles a rectangular
barrier.
Tunneling just below the top of such a barrier is very unprobable.
The thermally assisted tunneling is thus suppressed, and the thermal activation competes
directly with the ground-state tunneling, leading to a sharp escape-rate transition.

Mapping of a spin problem onto a particle one is, however, not a regular
procedure, and its form strongly depends on the form of the spin Hamiltonian
(see Ref. \onlinecite{zas90pla}). 
In many cases, such as for the model with $S_z^4$, it is difficult to find a
particle mapping.
On the other hand, one needs a simple criterion for roughly estimating the order
of the transition for different spin systems.
This criterion can be constructed as follows.
The first-order transitions occur if a spin system is close to a uniaxial
one,  ${\cal H} = F(S_z)$.
In this case the maximum of the classical energy of the spin, $U({\bf S})$, corresponds to a
certain value of the polar angle, $\theta=\theta_c$, and there is no saddle point of $U(\theta,\varphi)$. 
If nonuniaxial terms, such as the transverse field, are added to the
Hamiltonian, then a saddle point appears.
This brings the spin system closer to the common case, and if the saddle is
strongly pronounced, one can expect a second-order transition.
Thus, one can propose the following heuristic criterion for the boundary between first- and second-order transitions:
{\em the depth of the saddle is of the order of the height of the barrier}, i.e.,
$U_{\rm max}-U_{\rm sad} \sim U_{\rm sad}-U_{\rm min}$.
For the model with the transverse field one has 
$\Delta U \equiv U_{\rm sad}-U_{\rm min} = S^2 D (1-h_x)^2$ and 
$U_{\rm max}-U_{\rm sad} = 4S^2 D h_x$.
Equating these expressions one obtains $h_{xc} = 3-2\sqrt{2} \approx 0.17$,
which has a proper order of magnitude. 
In the case $H_z\ne 0$, the parameters of the barrier cannot be calculated
analytically, but estimations can be done.
In particular, in the strongly-biased case, $\delta \equiv 1-h_z \ll 1$, for
$h_x=0$ the top of the barrier corresponds to $\tilde\theta_c \equiv
\pi-\theta_c \cong \sqrt{2\delta}$, and $\Delta U \cong S^2 D \delta^2$.
For $h_x \ne 0$, one can use $U_{\rm max}-U_{\rm sad} \sim S^2 D h_x \tilde \theta_c$
and the same barrier height for estimations, which results in the correct dependence
$h_{xc} \sim \delta^{3/2}$ (Ref. \onlinecite{garmarchu98}).

In this communication we will study the quantum-classical escape-rate
transition for the spin model with a biaxial anisotropy, which is described by the Hamiltonian
%\marginpar{biaxham}
%
\begin{equation}\label{biaxham}
{\cal H} = -DS_z^2 - H_zS_z + B S_x^2 . 
\end{equation}
This model has the easy axis $z$, hard axis $x$ and thus the middle axis $y$.
Applying the longitudinal field is the simplest way to reduce the barrier.
Sometimes another form of the biaxial spin model, $ {\cal H} = K(
S_z^2 + \lambda S_y^2) + \cdots$,  is used, which for $\lambda < 1$ has the easy axis $x$, hard
axis $z$ and the middle axis $y$.
These two forms are related by $D=\lambda K$ and $B=(1-\lambda)K$.
A realization of this spin model is the magnetic molecule Fe$_8$ ($S=10$,
$D=0.31$~K, $B=0.092$~K, Refs. \onlinecite{baretal96} and \onlinecite{sanetal97}).

In the unbiased case, $H_z=0$, using the criterion 
$U_{\rm max}-U_{\rm sad} = U_{\rm sad}-U_{\rm min}$  readily yields $b_c =1$ for the
critical value of the reduced quantity $b\equiv B/D$ corresponding to the boundary
between the first- and second-order transitions.
Occasionally, this coincides with the exact value of $b_c$ which has been
obtained recently in Ref. \onlinecite{liamueparzim98} by the periodic
instanton method ($\lambda_c=1/2$). 
In the biased case, for the reduced energy $\tilde U \equiv U/(S^2D)$ one finds 
$\tilde U_{\rm min}=-1 +2h_z$, $\tilde U_{\rm sad} = h_z^2$, and $\tilde U_{\rm max} = b +h_z^2/(1+b)$.
Our criterion yields $b_c = (1-h_z)(\sqrt{1+h_z^2}-h_z)$, i.e., $b_c$
decreases linearly with $\delta \equiv 1-h_z$ in the strongly biased case.
Thus, in this case the region where a first-order transition can be expected, is wider for
the biaxial model than for the model with a transverse field ($h_{xc} \sim \delta^{3/2}$).

Now we proceed to an actual calculation of the boundary line $b_c(h_z)$ for
the model above.
In principle, this could be done by a kind of instanton or other
quasiclassical method.
However, the problem of obtaining relevant instantons 
(or even the action without computing instantons explicitly)  in the whole field range
appears to be mathematically difficult. 
Thus we choose to solve the problem with the help of the high-order
perturbation theory with respect to $b\equiv B/D$ in Eq. (\ref{biaxham}).
The latter was applied to calculate the splittings of the ground and excited
states for the unbiased model in Ref. \onlinecite{gar91jpa}.
It was shown that the instanton results of Ref. \onlinecite{enzsch86} for the
ground-state splitting are recovered in the limit $b\ll 1$, and that the Kramers degeneracy
for half-integer values of $S$ appears in a natural way.
For the model with a transverse field, this method was used in
Ref. \onlinecite{korshe78}
to calculate ground-state splitting in rare-earth compounds.
It has been generalized for excited states in Ref. \onlinecite{gar91jpa}.
Similar results have been obtained in Ref. \onlinecite{aubflaklaolb96} for
the model of two coupled nonlinear oscillators (quantum dimer), which can be
considered as a Schwinger-boson equivalent of the spin model with a transverse field.

The perturbative method was first to show the first- and second-order
transitions in the spin model with a transverse field in
Ref. \onlinecite{garchu97}.
The perturbatively determined value $h_{xc}=0.145$ was, however, wrong by a
numerical factor in comparison to the exact value $h_{xc}=1/4$ obtained subsequently by a
quasiclassical method. \cite{chugar97} 
The reason for this inaccuracy is that the order of the escape-rate transition
is controlled by the situation near the top of the barrier, just where the
perturbation theory breaks down.
A plausible way to improve the accuracy of the perturbative results is to
rescale them by fitting to the exact quasiclassical results at $h_z=0$.
Then one can expect improving the results in the whole range $0 < h_z < 1$,
where, for the biaxial model, exact quasiclassical calculations have not
yet been performed.  
At first we will test this method on the model with a transverse field, for which
the quasiclassical boundary line $h_{xc}(h_z)$ is known from 
Ref. \onlinecite{garmarchu98}.
We will see that, surprisingly, a simple rescaling of the perturbative results
yields the accurate quasiclassical ones in the whole range $0 < h_z < 1$.

Within the perturbative method, the quantum-classical transitions correspond
to different types of behavior of the function \cite{garchu97}
%\marginpar{Fpert}
%
\begin{equation}\label{Fpert}
f(m) = (\Delta\varepsilon_{mm'})^2 \exp(-\varepsilon_m/T) ,
\end{equation}
where $\varepsilon_m = -Dm^2 - H_z m$ are unperturbed energy levels of the
spin system and $\Delta\varepsilon_{mm'}$ is the splitting of the pair of in-resonance
levels $\varepsilon_m$ and $\varepsilon_{m'}$ belonging to different potential
wells.
This resonance condition is fulfilled for the values of the bias field 
$H_z = H_{zk} = kD$, where  $k=0, \pm 1, \pm 2, \ldots$; then one has  
$m' = -m-k$.
The function $f(m)$ is the quantum-mechanical tunneling probability 
$\Gamma_{mm'} \propto (\Delta\varepsilon_{mm'})^2$ weighed with the Boltzmann
exponent $\exp(-\varepsilon_m/T)$ characterizing the probability of thermal
activation to the level $m$.
The task is to find the escape level $m_{\rm esc}$, i.e., the value of $m$
which maximizes $f$.
The maximum of $f$ is searched at the interval $-S \leq m \leq m_b$, where
$m=-S$ corresponds to the bottom of the well and $m=m_b$ corresponds to the
top of the barrier which is determined from the condition that the level
splitting $\Delta\varepsilon_{mm'}$ reaches the value of the level separation 
$\varepsilon_m - \varepsilon_{m-1}$ (see, e.g., Ref. \onlinecite{garchu97}).
For temperatures above the transition temperature $T_0$, the result will be 
$m_{\rm esc}=m_b$, which corresponds to thermal activation over the barrier.
For $T<T_0$, one obtains $m_{\rm esc}(T) < m_b$, which means thermally
assisted tunneling (or ground-state tunneling for $m_{\rm esc} = -S$).
If the dependence $m_{\rm esc}(T)$ is smooth, the spin system undergoes a
second-order escape-rate transition.
If there is a jump of $m_{\rm esc}(T)$ at some temperature, there is a
first-order transition.

\begin{figure}[t]
\unitlength1cm
\begin{picture}(11,6.5)
\centerline{\epsfig{file=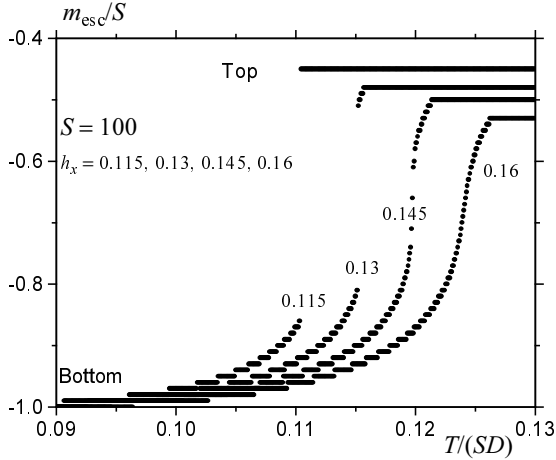,angle=-90,width=8cm}}
\end{picture}
%\par
%
\caption{ \label{tatq_meh}
The level making the dominant contribution into the escape rate 
$m_{\rm esc}$ vs temperature for the unbiased transversed field model.
}
\end{figure}
\begin{figure}[t]
\unitlength1cm
\begin{picture}(11,6.5)
\centerline{\epsfig{file=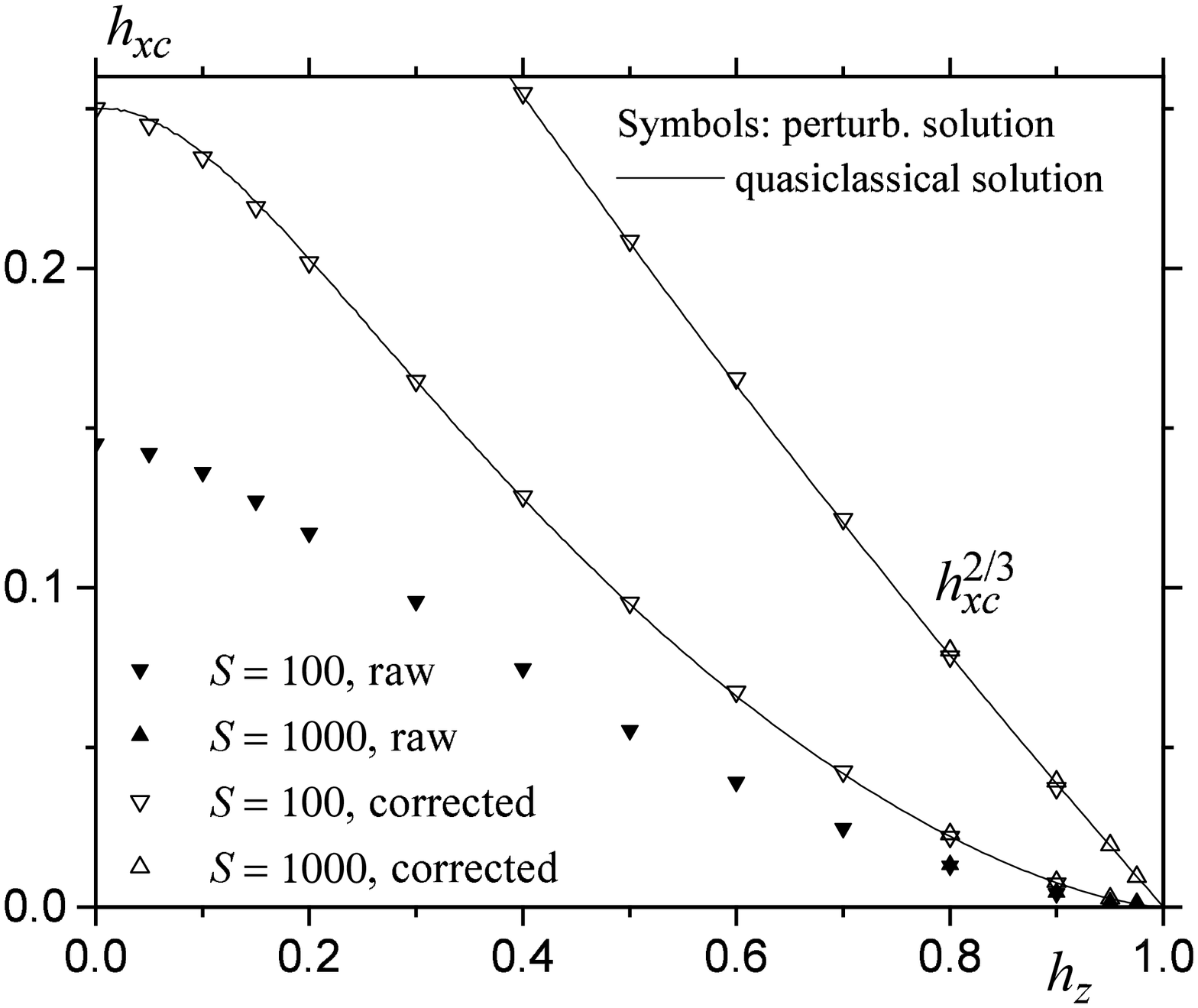,angle=-90,width=8cm}}
\end{picture}
%\par
%
\caption{ \label{tatq_hxc}
Boundary between the first- and the second-order transitions for the model
 ${\cal H} = -DS_z^2 -H_zS_z - H_x S_x$.
}
\end{figure}
\begin{figure}[t]
\unitlength1cm
\begin{picture}(11,6.5)
\centerline{\epsfig{file=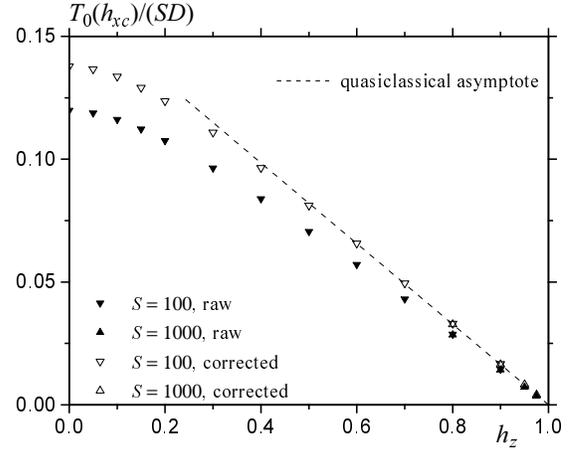,angle=-90,width=8cm}}
\end{picture}
%\par
%
\caption{ \label{tatq_t0h}
Quantum-classical transition temperature $T_0$ 
at the boundary between first- and second-order transitions for the model
 ${\cal H} = -DS_z^2 -H_zS_z - H_x S_x$.
}
\end{figure}

For the model with an arbitrarily directed field and without transverce
anisotropy the level splittings are given by the formula 
\begin{eqnarray}\label{split}
&&
\Delta \varepsilon_{mm'} = \frac{2D}{[(m'-m-1)!]^2}
\nonumber\\
&&
\qquad
\times \sqrt{\frac{ (S+m')! (S-m)! }{ (S-m')! (S+m)! } }
\left( \frac{ H_x }{ 2D } \right)^{m'-m},
\end{eqnarray}
which is a generalization of the zero-bias result of 
Ref. \onlinecite{gar91jpa} (see, e.g., Ref. \onlinecite{garchu97}).
Numerical maximization of the function $f(m)$ of Eq. (\ref{Fpert}) leads to the
results for $m_{\rm esc}(T)$ shown in Fig. \ref{tatq_meh}.
One can see that the transition is second order for $h_x > 0.145$ and first
order for $h_x < 0.145$.
The results in Fig. \ref{tatq_meh} suggest that in some range of $h_x$ (as,
e.g., for $h_x=0.13$) there are two transition temperatures: With lowering
temperature, $m_{\rm esc}$ at first deviates from $m_b$ (a second-order
transition) and then jumps downwards (a first-order transition).
In fact, this feature is an artifact of the perturbative approach to
tunneling, which breaks down near the top of the barrier.
(For this reason also the crossover field $h_{xc}=0.145$ substantially deviates from the exact
value $h_{xc}=0.25$ obtained in Ref. \onlinecite{chugar97}.)
In the following we will associate the crossover from first- to
second-order transitions with the transverse field at which the jump in the
dependence $m_{\rm esc}(T)$ appears.

The values of the crossover field $h_{xc}$ for different
longitudinal fields $h_z$, which have been obtained by the perturbative method
described above, are shown by the solid symbols in Fig. \ref{tatq_hxc}.
These data can be corrected by multiplying by the factor 0.25/0.145=1.274 to
fit to the exact quasiclassical value $h_{xc}=0.25$ in the unbiased case.
Surprizingly, this rescaling leads to the results which completely coincide
in the whole range of the bias field with the results obtained in Ref. \onlinecite{garmarchu98} 
by the quasiclassical method based on the particle mapping.

The results for the transition temperature $T_0$ at the boundary between
first- and second-order escape-rate transitions in the whole range of the bias
field is shown in Fig. \ref{tatq_t0h}.
Again, the perturbative results can be corrected by multiplying by
0.1378/0.1198=1.150 to fit to the exact value in the unbiased case: 
$T_0(h_{xc}) = [\sqrt{3}/(4\pi)] SD = 0.1378\, SD$ (Ref. \onlinecite{chugar97}).
This makes them accurate in the whole range of $h_z$, as shows the comparison
with the exact asymptote $T_0(h_{xc}) \cong 0.1642\, (1-h_z) SD$ 
(Ref. \onlinecite{garmarchu98}) in the strongly biased limit.

\begin{figure}[t]
\unitlength1cm
\begin{picture}(11,6.5)
\centerline{\epsfig{file=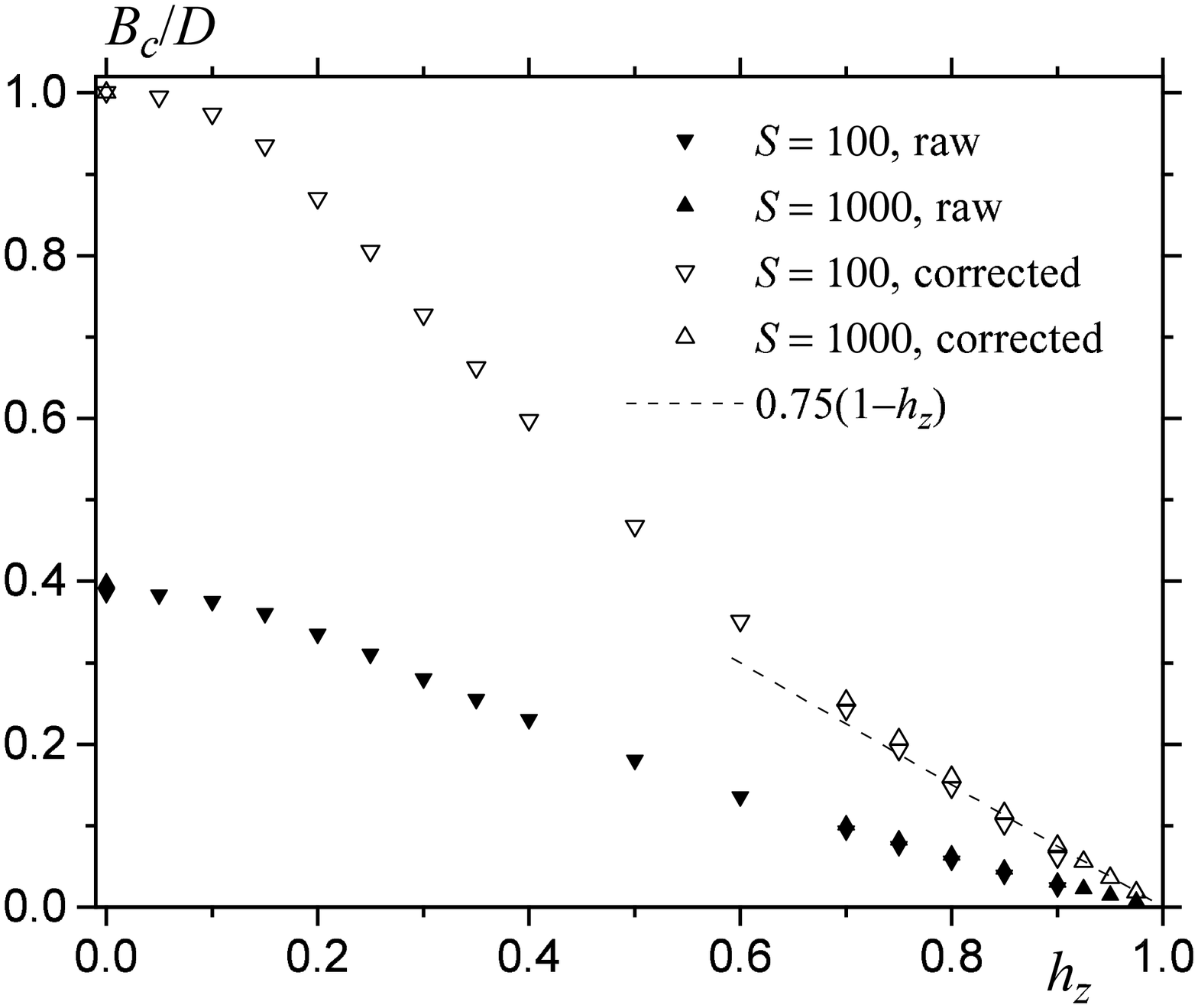,angle=-90,width=8cm}}
\end{picture}
%\par
%
\caption{ \label{tatq_bc}
Boundary between the first- and the second-order transitions for the model
 ${\cal H} = -DS_z^2 -H_zS_z + B S_x^2$.
}
\end{figure}
\begin{figure}[t]
\unitlength1cm
\begin{picture}(11,6.5)
\centerline{\epsfig{file=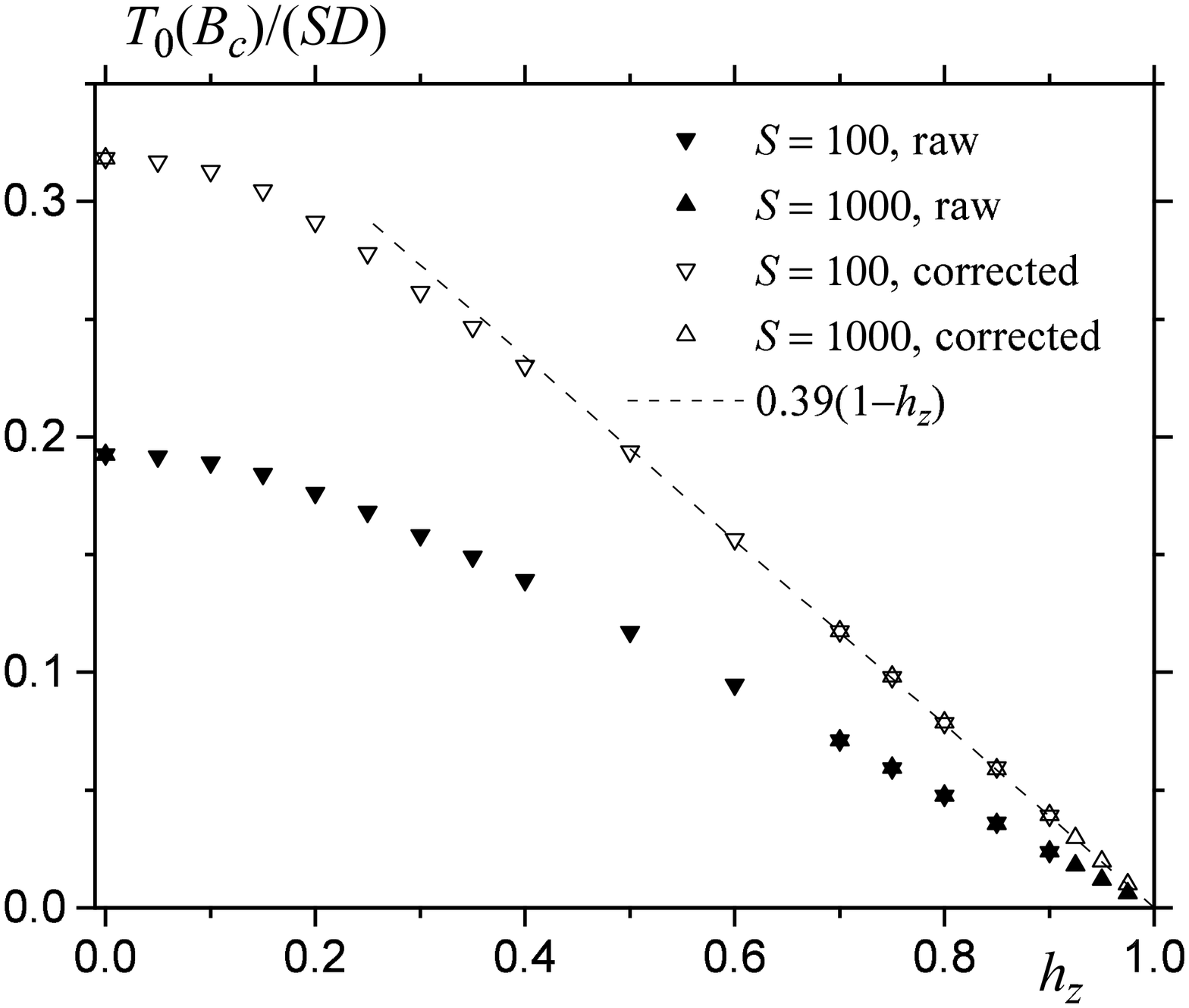,angle=-90,width=8cm}}
\end{picture}
%\par
%
\caption{ \label{tatq_t0b}
Quantum-classical transition temperature $T_0$ 
at the boundary between first- and second-order transitions for the model
 ${\cal H} = -DS_z^2 -H_zS_z + B S_x^2$.
}
\end{figure}

Having tested the perturbative method with rescaling on the model with an
arbitrarily directed field, let us now proceed to the biaxial model of
Eq. \ref{biaxham}, where a quasiclassical solution is still lacking in the
biased case, $H_z\ne 0$.
For this model the formula for the level splittings reads
%\marginpar{splitbiax}
%
\begin{eqnarray}\label{splitbiax}
&&
\Delta \varepsilon_{mm'} = \frac{ 8D }{ \{[(m'-m)/2-1]!\}^2 }
\nonumber\\
&&
\qquad
\times \sqrt{\frac{ (S+m')! (S-m)! }{ (S-m')! (S+m)! } }
\left( \frac{ B }{ 16D } \right)^{(m'-m)/2},
\end{eqnarray}
which is a generalization of the zero-bias result of Ref. \onlinecite{gar91jpa}.
This formula is only applicable if $m'-m$ is even, otherwise the
perturbative method yields $\Delta \varepsilon_{mm'}=0$, i.e., the Kramers degeneracy.
Hence, for $S$ integer and the resonant values of the
bias field $H_z = H_{zk} = kD$, the level pairs are splitted for  $k=0, \pm 2, \pm 4, \ldots$,
but remain degenerate for $k=\pm 1, \pm 3, \ldots$.
For $S$ half integer, the situation is reversed.
Below we will only consider the resonant values of $H_z$ for which there is no
Kramers degeneracy.

The dependences $m_{\rm esc}(T)$ for the biaxial model are similar to those
for the model with an arbitrarily directed field, and they are assessed in the
same way.
The results for the bondary between first- and second-order transitions,
$B_c$, are shown in Fig. \ref{tatq_bc}.
For rescaling of these results we have used the exact value $B_c/D =1$ at
$H_z=0$ (Ref. \onlinecite{liamueparzim98}).
One can see that $B_c \propto 1-h_z$ in the strongly biased case, as was
conjectured above.
The transition temperature $T_0$ at the boundary between
first- and second-order transitions in the whole range of the bias
field is shown in Fig. \ref{tatq_t0b}.
Here we have used for the rescaling of the results the exact value $T_0(B_c) =
(1/\pi) SD = 0.3183\, SD$ at $H_z=0$, which can be extracted from Ref. \onlinecite{liamueparzim98}.

In comparison to the previously considered model with an arbitrarily directed
field, here the crossover between first- and second-order transitions occurs
at higher values of the control parameter $B$, thus the range of the
first-order transitions is extended.
On the other hand, for this reason using a perturbation theory in $B/D$ is
less justified, and one can expect some deviations of the rescaled perturbative results
from the exact quasiclassical ones for $H_z\ne 0$.
Thus obtaining of the latter is an actual problem.

For single-domain magnetic particles with small transverse anisotropy $B$,
i.e., for nearly uniaxial ones, the crossover from first- to second-order
escape-rate transitions should be searched for in the strongly biased case,
$\delta \equiv 1-h_z \ll 1$, where the barrier is reduced.
Here our result for the transition temperature at the boundary between first-
and second-order transitions is $T_0(B_c) \approx 0.39\,SD\delta$
(see Fig. \ref{tatq_t0b}).
On the other hand, the unperturbed barrier height is given by 
$\Delta U = S^2 D \delta^2$.
Eliminating $\delta$ from the above formulas, one can relate $T_0$ and
$\Delta U$ as follows: $T_0 \approx (0.39)^2 D \Delta U/T_0$.
The ratio $\Delta U/T_0$ is fixed by the requirement that the escape rate be
not too high and not too low.
Adopting a typical value  $\Delta U/T_0=40$, as was done in Ref. \onlinecite{garmarchu98} 
for the model with an arbitrarily directed field, one arrives at the
estimation $T_0(B_c) \approx 6.3 D$, which is substantially higher than that of
Ref. \onlinecite{garmarchu98}, $T_0(h_{xc}) \approx D$.

This work has been supported in part by the U.S.
National Science Foundation through Grant No. DMR-9024250.
D. G. thanks S. Flach for pointing out the analogy with the quantum dimer
problem and for the critical reading of the manuscript.

%\bibliography{gar}

\end{document}